\documentclass[conference]{IEEEtran}
\usepackage{cite}
\usepackage{amsmath,amssymb,amsfonts}
\usepackage{graphicx}
\usepackage{textcomp}
\usepackage{xcolor}
\usepackage{fancyhdr}
\usepackage[hyphens]{url}
\usepackage[noend]{algpseudocode}
\usepackage{algorithm}
\usepackage{enumitem}
\usepackage{subfig}
\graphicspath{ {./figures/} }
\newcommand*{\thead}[1]{\multicolumn{1}{|c|}{\bfseries #1}}
\def\BibTeX{{\rm B\kern-.05em{\sc i\kern-.025em b}\kern-.08em
    T\kern-.1667em\lower.7ex\hbox{E}\kern-.125emX}}

\IEEEoverridecommandlockouts
\begin{document}
%%%%%%%%%%%---SETME-----%%%%%%%%%%%%%
\title{\LARGE Prediction-Based Power Oversubscription in Cloud Platforms
 %\vspace{-.54in}
}
\author{
  Alok Kumbhare, Reza Azimi, Ioannis Manousakis, Anand Bonde, Felipe Frujeri, Nithish Mahalingam, \\
  Pulkit Misra, Seyyed Ahmad Javadi, Bianca Schroeder, Marcus Fontoura, and Ricardo Bianchini \\
\\
  Microsoft Azure and Microsoft Research\footnotemark\textsuperscript{*}\thanks{\textsuperscript{*}Azimi, Javadi, and Schroeder were at Microsoft Research during this work.}
}
%%%%%%%%%%%%%%%%%%%%%%%%%%%%%%%%%%%%
\maketitle
\pagestyle{plain}

%%%%%% -- PAPER CONTENT STARTS-- %%%%%%%%

\begin{abstract}
Datacenter designers rely on conservative estimates of IT equipment
power draw to provision resources.
This leaves resources underutilized and requires more datacenters to
be built.  Prior work has used power capping to shave the rare power
peaks and add more servers to the datacenter, thereby oversubscribing
its resources and lowering capital costs. This works well
when the workloads and their server placements are known.
Unfortunately, these factors are unknown in public clouds, forcing
providers to limit the oversubscription so that performance is never
impacted.

In this paper, we argue that providers can use predictions of workload
performance criticality and virtual machine (VM) resource utilization
to increase oversubscription. This poses many challenges, such as
identifying the performance-critical workloads from black-box VMs,
creating support for criticality-aware power management, and
increasing oversubscription while limiting the impact of capping.
We address these challenges for the hardware and software
infrastructures of Microsoft Azure. The results show
that we enable a $2\times$ increase in oversubscription
with minimum impact to critical workloads.
\end{abstract}

\section{Introduction}
\label{sec:intro}

\noindent{\bf Motivation.} Large Internet companies continue building
datacenters to meet the increasing demand for their services. Each
datacenter costs hundreds of millions of dollars to build. Power plays
a key role in datacenter design, build out, IT capacity deployment,
and physical infrastructure cost.

The power delivery infrastructure forms a hierarchy of devices that
supply power to different subsets of the deployed IT capacity at the
bottom level. Each device includes a circuit breaker to prevent damage
to the IT infrastructure in the event of a power overdraw. When a
breaker trips, the hardware downstream loses power, causing a partial
blackout.

To avoid tripping breakers, designers conservatively
provision power for each server based on either its maximum nameplate
power or its peak draw while running a power-hungry benchmark, such as
SPEC Power~\cite{Lange2009}. The maximum number of servers is then the
available power (or breaker limit) divided by the per-server
provisioned value. This provisioning leads to massive power under-utilization.
As the IT demand increases, it also requires building new datacenters
even when there are available resources (space, cooling, networking)
in existing ones, thus incurring huge unnecessary capital costs.

To improve efficiency and avoid these costs, prior work has proposed
combining power capping and
oversubscription~\cite{Fan2007,Wu2016}. The idea is to leverage actual
server utilization and statistical multiplexing across workloads to
oversubscribe the delivery infrastructure by {\em adding more servers
  to the datacenter,} while ensuring that the power draw remains below
the breakers' limits. This is achieved by continuously monitoring the
power draw at each level and using power capping (via CPU
voltage/frequency and memory bandwidth throttling), when necessary.
As throttling impacts performance, these approaches carefully define
which workloads can be throttled and by how much. For example,
Facebook's Dynamo relies on predefined workload priority groups, and
throttles each server based on the priority of the workload it
runs~\cite{Wu2016}. Using Dynamo, Facebook was able to oversubscribe
its datacenters by 8\%, lowering costs by hundreds of millions of
dollars.

This oversubscription approach works well when workloads and their
server placements are known. Unfortunately, {\em public cloud
  platforms violate these assumptions.} First, each server runs many
VMs, each with its workload, performance, and power characteristics.
Hence, throttling the entire server would impact performance-critical
(e.g., interactive services) and non-critical (e.g., batch) workloads
alike. Second, VMs dynamically arrive and depart from each server,
producing varying mixes of characteristics and preventing predefined
server groupings or priorities. Third, each VM must be treated as a
black box, as customers are often reluctant to accept deep inspection
of their VMs.  Thus, the platform does not know which VMs are
performance-critical and which ones are not. For these reasons, {\em
  oversubscription in public clouds has been limited so that
  performance is never impacted.}

\noindent{\bf Our work.} In this paper, we argue that cloud providers
can increase oversubscription substantially by carefully scheduling
VMs and managing power, based on predictions of workload performance
criticality and VM CPU utilization.  Our insight is that there are
many non-critical workloads (e.g., batch jobs) that can tolerate a
slightly higher rate of capping events and/or deeper throttling; the
capping of performance-critical workloads must be controlled more
tightly.  Using predictions to identify these workloads and place them
carefully across the datacenter provides the power slack and
criticality-awareness needed to increase oversubscription.

Accurately predicting (black-box) workload criticality is itself a
challenge.  Prior work~\cite{Cortez2017} associated a diurnal
utilization pattern with user interactivity and the critical need for
high performance.  It inferred criticality using the Fast Fourier
Transform (FFT) algorithm on the CPU utilization time series for each
workload.  Here, we present a more accurate and robust
pattern-matching algorithm, and a machine learning (ML) model that
uses the algorithm's output during training. We also propose a model
for predicting the 95th-percentile CPU utilization over a VM's
lifetime.

With these predictions, we increase the power slack in the datacenter
by balancing the expected power draw and our ability to lower it via
throttling when a power budget is exceeded (causing a capping
event). We accomplish this with a criticality- and utilization-aware
VM placement policy. When events occur, we must cap power
intelligently as well. So, we propose a system that protects
performance-critical VMs from throttling when capping a server's power
draw.

Using the above contributions and the history of power draws, we
devise a new strategy for selecting the amount of
oversubscription. The strategy limits the impact of capping on the two
VM types to predefined acceptable values, thereby enabling significant
but controlled increases in oversubscription.  {\em Providers that
  prefer to treat all external (i.e., third-party) VMs the same can
  simply assume them all to be critical, and classify only the
  internal (i.e., first-party) VMs into the two types at the cost of
  a lower increase in oversubscription.}

We implement our work 
for the hardware, firmware, and software of Microsoft Azure.
Specifically, we implement our criticality algorithm and ML models for
Azure's ML and prediction-serving system; implement our VM
placement policy as an extension of Azure's VM scheduler;
implement our per-VM capping system as a host-level agent on each
server, and leverage Azure's capping mechanisms; and implement
our oversubscription strategy using Azure's historical power
and utilization telemetry.

The evaluation shows that our criticality algorithm is substantially
more accurate than FFTs, and that our ML models achieve high
prediction accuracy. Our system and policy have not yet been deployed
widely, so we evaluate them on the same hardware, firmware, and
software that runs in production, but in an isolated environment. We
also show that our system and policy lower the performance impact of a
capping event, while our policy produces fewer events.
Overall, we can increase oversubscription by $2\times$ (from 6 to
12\%), compared to the state-of-the-art approach.
This increase would save \$75.5M in capital costs from just one
datacenter site (128MW).  Assuming that all external VMs are
performance-critical would lower the savings to a still significant
\$28.1M.  Large cloud providers have many such sites, so these savings
would multiply.

We have started deploying our work in production at Microsoft
  and mention some lessons in Section~\ref{sec:lessons}.

\noindent{\bf Related work.} The prior work on power
capping~\cite{Guliani19,Isci06,Lefurgy07,Liu16,Lo15,Ma11,Mishra10,
  Raghavendra2008,Zhang16} and
oversubscription~\cite{Fan2007,govindan2009,Hsu2018,Li19,Sakamoto17,
  Wang2016,Wu2016} produced major advances in server and datacenter
power management. Unfortunately, it falls short for real public
clouds. For example, it has capped server power using inputs that are
typically not available in the public cloud, such as application-level
metrics or operator annotations. Moreover, it has employed reactive
and expensive VM migration in clusters, instead of leveraging
predictions for capping and capping-aware scheduling. Prior
oversubscription works have focused on non-cloud datacenters and
full-server capping when workloads and their priorities are known.

\noindent{\bf Summary.} We make the following main contributions:
\begin{enumerate}[wide,labelwidth=!,labelindent=10pt,topsep=0pt,itemsep=-1ex,partopsep=1ex,parsep=1ex]
\item An algorithm and ML model for predicting performance criticality, and a model for predicting VM utilization.
\item A VM placement policy that uses these predictions to minimize the number of capping events and their impact.
\item A per-VM power capping system that uses predictions of criticality to protect certain VMs.
\item A strategy that leverages the contributions above to increase the amount of oversubscription.
\item Implementation and results for the infrastructure of Azure, showing large potential cost savings.
\item Lessons from production deployment of our work.
\end{enumerate}  

Though we build upon Azure's infrastructure, our conceptual
contributions (e.g., predicting criticality; using predictions in VM
placement, power budget enforcement, and oversubscription) apply
directly to any cloud platform.
\section{Background and Context}
\label{sec:background}

\subsection{Typical power delivery and server deployment}

At the top of the power delivery hierarchy, the electrical grid
provides power to a sub-station that is backed up by a generator.  An
Uninterruptible Power Supply (UPS) unit provides battery backup
while the generator is starting up.
The UPS feeds one power distribution unit (PDU) per row of servers.
Each row PDU supplies power to several rack PDUs, each of which
feeds a few server chassis. Each chassis contains a few power
supplies (PSUs) and dozens of blade servers.\footnote{We refer to
  ``blades'' and ``servers'' interchangeably throughout the paper.}  A
PDU trips its circuit breaker when the power draw exceeds the rated
value (budget) for the unit, causing a power outage.

Designers deploy servers so that breakers never trip, leading to
wasted resources (chiefly space, cooling, and networking). Combining
hierarchical power capping and oversubscription enables more capacity
to be deployed and better utilizes
resources~\cite{Fan2007,Wu2016}. For example, if designers find that
the historical per-row power draw is consistently lower than the row
PDU budget, they can ``borrow'' power from each row to add more rows
(until they run out of row space) under the
UPS budget. The extra rows oversubscribe the power at the
UPS level.
The servers downstream from the oversubscribed PDUs/UPS must be
power-capped, whenever they are about to draw power that has already
been borrowed.

\begin{figure}
\includegraphics[width=\columnwidth]{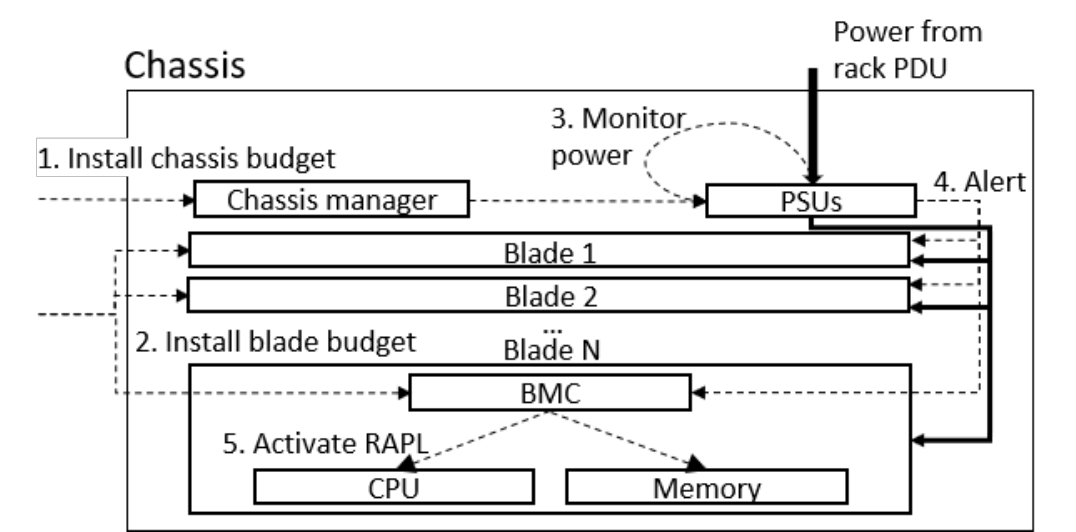}
\vspace{-.2in}
\caption{Chassis-level alert-based power capping.}
\label{fig:chassispowercap}
\vspace{-.2in}
\end{figure}

\subsection{Azure's existing power capping mechanisms}

For clarity and ease of experimentation, in this paper we
  explore power budget enforcement at the chassis level.

Figure \ref{fig:chassispowercap} shows Azure's current
mechanisms.  Each chassis contains a manager that exposes the
management interface. To enable capping, Azure first sets the
total chassis-level power budget at the chassis manager and PSUs (step
1), and each individual blade power cap in its board management chip
(BMC) (step 2). Azure sets each blade budget to its even share
of the chassis budget.

Under normal operation, no capping takes place, i.e. each server is
free to draw more power than its even share, as long as the total
chassis draw is below the chassis budget.  The PSUs monitor the
chassis draw (step 3) and alert the blades directly when the chassis
budget is about to be exceeded (step 4).  Upon an alert, the blade
power must be brought below its even-share cap.  (Uneven caps are
infeasible due to the overhead of dynamically reapportioning and
reinstalling blade budgets.)
The BMC splits the cap evenly across its sockets and uses Intel's
Running Average Power Limit (RAPL)~\cite{RAPL} to lower the blade
power (step 5). RAPL throttles the entire socket (slowing down {\em
  all cores equally}) and memory using a feedback loop until the cap
is respected. Typically, RAPL brings the power below the cap in less
than 2 seconds.

\subsection{Azure's existing VM scheduler}

A cloud platform first routes an arriving VM to a 
server cluster. Within each cluster, a VM scheduler is responsible for
placing the VM on a server. The scheduler uses heuristics to tightly pack VMs,
considering the incoming VM's multiple resource requirements and each
server's available resources.

Azure's scheduler implements its heuristics as two sets of
rules.  It first applies {\em constraint} rules (e.g., does the server
have enough resources for the VM?) to filter invalid servers, and then
applies {\em preference} rules each of which orders the candidate
servers based on a preferred metric (e.g., a packing score derived
from available resources). It then weights each candidate based on its
order on the preference list for each rule and the rule's pre-defined
weight. Finally, it picks a server with the highest aggregate weight
for allocation.  No rules currently consider power draws or capping.

\subsection{Azure's existing ML system}

To integrate predictions into VM scheduling in practice, we
target Resource Central~\cite{Cortez2017},
the existing ML and prediction-serving system in Azure.
The system provides a REST service for clients (in
our case the VM scheduler) to query for predictions.
It can receive input
features (e.g., user, VM size, guest OS) that are known at deployment
time from the scheduler, execute one of our ML models (criticality or
CPU utilization), and respond with the prediction and a confidence
score.  Model training is done in the background, e.g. once a day.

\section{Prediction-Based Oversubscription}
\label{sec:solution}

Cloud providers provision servers
conservatively, and have full-server capping mechanisms and
capping-oblivious VM schedulers.  We propose to provision servers more
aggressively by making the infrastructure smarter and finer-grained
via VM behavior predictions.  Key challenges include having to create
or adapt many components (e.g., scheduler, chassis manager) to the
predictions, while ensuring that the tradeoff between oversubscription
and performance is tightly controlled.

Next, we overview our design and then detail its main components.
Then, we describe our strategy for provisioning servers to balance
cost savings and performance impact.

\subsection{Overview}

\begin{figure}
\centering
\includegraphics[height=1.75in,width=\columnwidth]{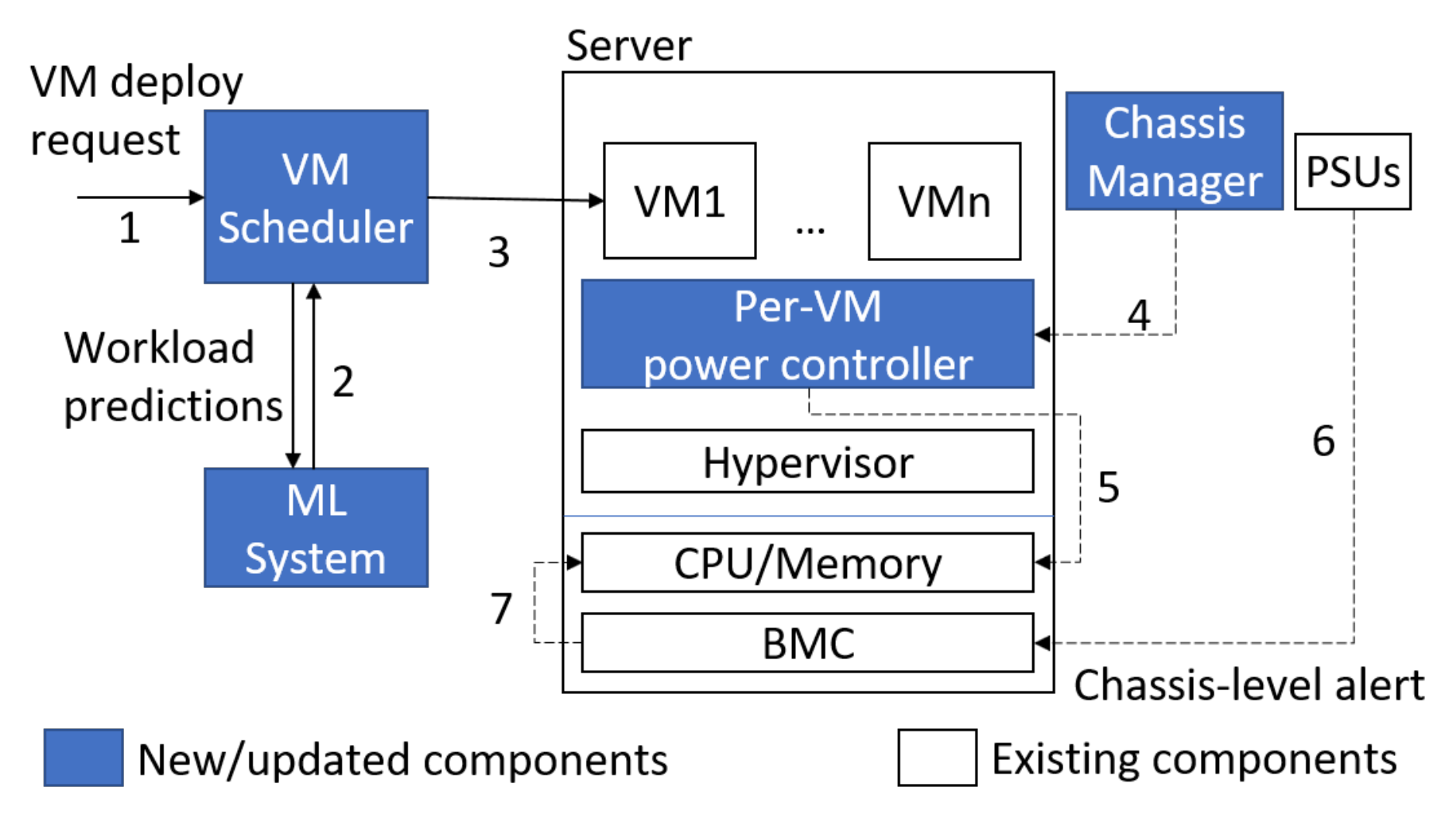}
\vspace{-.25in}
\caption{System overview.}
\label{fig:sysarch}
\vspace{-.25in}
\end{figure}

Figure~\ref{fig:sysarch} overviews our system and its operation,
showing the existing and new/modified components in different colors.
We modify the VM scheduler to use predictions of VM performance
criticality and resource utilization. We implement our ML models so
they can be managed and served by the existing ML and
prediction-serving system.  We modify the chassis manager to query the
chassis power draw and interact with our new per-VM power
controller. The controller manages its server's power draw during a
capping event.

In detail, a request to deploy a set of VMs arrives at our VM
scheduler (arrow \#1).  The scheduler then queries the ML system for
predictions of workload performance criticality and resource
utilization (\#2).  Using these predictions, it decides on which
servers to place the VMs (\#3).  After selecting each VM's placement,
the scheduler tags the VM with its predicted workload type and
instructs the destination server to create it.
Each chassis manager frequently polls its local PSUs to determine
whether the power draw for the chassis crosses a threshold just below
the chassis budget. (This threshold enables the controller to perform
per-VM capping and hopefully avoid needing full-server RAPL.) When
this is the case, the manager alerts the controller of each server in
the chassis (\#4).

Upon receiving the alert, the controller at each server manages the
server's (even) share of the chassis budget across the local VMs based
on their workload types.  It does this by first throttling the CPU
cores used by non-performance-critical VMs (\#5).  Throttling these
VMs may be enough to keep the power below the chassis budget, and
protect the performance-critical VMs.  If it is not enough, the PSUs
alert the servers' BMCs (\#6), which will then use RAPL as a last
resort to lower the chassis power below the budget (\#7).

\noindent{\bf Limiting impact on non-critical VMs.} Though we protect
critical VMs from throttling, we limit the performance impact on
non-critical VMs in three ways. First, for long-term server
provisioning, our oversubscription strategy carefully selects chassis
budgets to limit the number of capping events and their severity to
predefined acceptable values (e.g., no more than 1\% events
for non-critical VMs, each lowering the core frequency to no less than
75\% of the maximum). Second, for medium-term management, our
scheduler places VMs seeking to minimize the number of events and
their severity. Finally, in the shortest term, our per-VM controller
increases the core frequency of non-critical VMs as soon as possible.

\noindent{\bf Treating external VMs as performance-critical.}
Provi\-ders who prefer to treat all paying customers the same can
easily do so by assuming that all external VMs are
performance-critical; only internal VMs (e.g., running the
  provider's own managed services) would be classified into the two
criticality types.  We explore this assumption in
Section~\ref{sec:eval-oversub}.

\subsection{Predicting VM criticality and utilization}
\label{sec:ML}

Our approach depends on predicting VM performance criticality and
utilization at arrival time, i.e. just before the VMs are deployed.
We train supervised ML models to produce these predictions based on
historical VM arrival data and telemetry that was collected after
those VMs were deployed. Since VMs are black boxes, our telemetry
consists of CPU utilization data only, as deep inspection is not an
option.

\noindent{\bf Inferring criticality.}  Predicting criticality requires
a method to determine the VM {\em labels}, i.e. whether the workload
of each VM is performance-critical or not, before we can train
a model.  As in prior work~\cite{Cortez2017}, we consider a
workload critical if it is user-facing, i.e. a human is interacting
with the workload (e.g., front-end webservers, backend databases), and
less critical otherwise (e.g., batch, development and testing
workloads). As user-facing workloads exhibit utilization patterns that
repeat daily (e.g., high during the day, low at night), the problem
reduces to identifying VMs whose time series of CPU utilizations
exhibit 24-hour periods~\cite{Cortez2017}.

Obviously, some background VMs may exhibit 24-hour periods.  This is
not a problem as we seek to be conservative (i.e., it is fine to
classify a non-user-facing workload as user-facing, but not
vice-versa).  Moreover, some daily batch jobs have strict deadlines,
so classifying them as user-facing correctly reflects their needs.
Importantly, focusing on the CPU utilization signal works well even
when the CPU is not the dominant resource, as the CPU is always a good
proxy for periodicity (e.g., network-bound interactive workloads
exhibit more CPU activity during the day than at night).

We considered but discarded other approaches for inferring whether a
VM's workload is user-facing.  For example, observing whether a VM
exchanges messages does not work because many non-user-facing
workloads communicate externally (e.g., to bring data in for batch
processing).

\noindent{\bf Identifying periodicity.} There are statistical methods
for identifying periods in time series, such as FFT or the
auto-correlation function (ACF). For example, \cite{Cortez2017}
assumes a workload is user-facing if the FFT indicates a 24-hour
period.  {\em We evaluated ACF and FFT methods on 840 workloads on
Azure}.  Surprisingly, we find that both methods lead to frequent
mis-classifications.  We identify three culprits:
\begin{enumerate}[wide,labelwidth=!,labelindent=10pt,topsep=0pt,itemsep=-1ex,partopsep=1ex,parsep=1ex]
\item The diurnal patterns in user-facing workloads often have
  significant noise and interruptions. For example, we observe
  user-facing workloads with clear 24-hour periods for many days,
  interrupted by a period of constant or random load, causing them to
  be mis-classified as non-user-facing.

\item The diurnal patterns often exhibit increasing/decreasing trends
  (e.g., the workload becomes more popular over time), and varying
  magnitudes of peaks/valleys across days.  These effects cause some
  user-facing workloads to be mis-classified.

\item There are many machine-generated workloads with periods of 1
  hour, 4 hours, 6 hours or other divisors of 24 hours, which
  therefore also have 24-hour periods, leading to machine-generated
  workloads that are mis-classified as user-facing.
\end{enumerate}

Part of the problem is that ACF and FFT are very general tools with
different goals, e.g. decomposing a signal for compact representation
and capturing general correlations, not solutions for our specific
problem of 24-hour periods.

\noindent{\bf Criticality algorithm.} Thus, we devise a new algorithm
that is more robust and targeted at our specific problem.  Our idea is
to extract from a VM's utilization time series a {\em template} for a
typical 24-hour period and then check how well this template captures
most days in the series. We design the template extraction and
comparison to be robust to noise and interruptions to deal with issue
\#1 above. We pre-process the data using methods from time series
analysis
to address \#2. To
deal with \#3, we extract templates for shorter periods (8 and 12
hours) and ensure that the 24-hour template is the best fit. These
periods subsume the other short periods.

More precisely, the input to our pattern-matching algorithm is the
average CPU utilization for each 30-minute interval over 5
weekdays. (Shorter workloads cannot be classified and should be
conservatively assumed user-facing.)  For each utilization time
series, the algorithm does the following:
\begin{enumerate}[wide,labelwidth=!,labelindent=10pt,topsep=0pt,itemsep=-1ex,partopsep=1ex,parsep=1ex]
\item It de-trends and normalizes the time series, so that all days
  exhibit utilizations within the same rough range.  De-trending
  scales each utilization based on the mean of the previous 24 hours,
  whereas normalization divides each utilization by the standard
  deviation of the whole time series.
  
\item It extracts the 24-hour template by identifying, for each time
  of the day (in 30-minute chunks), its ``typical'' utilization
  computed as the median of all utilizations in the pre-processed
  series that were reported at this time of the day.

\item It overlays the template over the pre-processed series for each
  day and computes the average deviation for
  each utilization, after excluding the 20\% largest deviations.
  
\item It repeats steps 2 and 3 to compute average deviations for
  8-hour and 12-hour templates, and then computes two scores: 24-hour
  average deviation divided by 8-hour average deviation (called {\tt
    Compare8}), and 24-hour average deviation divided by 12-hour
  average deviation (called {\tt Compare12}). If the scores are close
  to 0, the workload is likely to be user-facing.  Ultimately, it
  classifies a time series as user-facing, if its Compare8 value is
  lower than a threshold (Section~\ref{sec:alg-models}).
\end{enumerate}

\noindent {\bf Criticality prediction.} The algorithm above produces
labels that we use to train an ML model to classify arriving VMs as
user-facing or non-user-facing.  Specifically, we train a Random
Forest using the labels and many features (pertaining to the arriving
VM and its cloud subscription) available at arrival time: the
percentage of user-facing VMs in the subscription, the percentage of
VMs that lived at least 7 days in the subscription, the total number
of VMs in the subscription, the percentage of VMs in each CPU
utilization bucket, the averages of the VMs' average and
95th-percentile CPU utilizations in the subscription, the arriving
VM's number of cores and memory size, and the arriving VM's type.

\noindent{\bf Utilization prediction.} For utilization predictions, we
train a two-stage model to predict 95th-percentile VM CPU utilization
based on labels produced by previous VM executions (actual
95th-percentile utilizations over the VMs' lifetimes) and the same VM
features we use in the criticality model.  Since predicting
utilization exactly is hard, our model predicts it into 4 buckets:
0\%-25\%, 26\%-50\%, and so on.  The first stage of the model is a
Random Forest that predicts whether or not the 95th-percentile
utilization is above 50\%.  In the second stage, we have a Random
Forest for buckets 1-2 and another for buckets 3-4.  We train these
latter forests with just the VMs we can predict with high-confidence
($\ge$ 60\%) in the first stage.

We experimented with single-stage models, but they did not produce
accurate predictions with enough confidence.

\subsection{Modified VM scheduler}

Our ability to increase oversubscription and the efficacy of the
per-VM power controller depend on the placement of VMs in each
cluster. Better placements have a balanced distribution of power draws
across the different chassis to reduce the number of capping events
(Goal \#1); and a balanced distribution of cap-able power (drawn by
non-user-facing VM cores) across servers, so the controller can lower
the power during an event without affecting critical VMs (Goal
\#2). The scheduler must remain effective at packing VMs while
minimizing the number of deployment failures (Goal \#3).

Given these goals, we modify Azure's VM scheduler to
become criticality- and utilization-aware, using predictions at VM
arrival time.
Our policy is a preference rule that sorts the feasible servers based
on a ``score''.  Each server's score considers the predicted
95th-percentile CPU utilization of the VMs already placed in the same
chassis (targets Goal \#1), and the predicted criticality and
95th-percentile CPU utilization of the VMs already placed on the same
server (targets Goal \#2).
The policy only considers CPU utilization because the CPUs are
 the dominant source of dynamic power in Azure's servers.
As Section~\ref{sec:scheduler} shows, our policy does not degrade
the packing of VMs onto servers, nor does it increase the percentage
of VM deployment failures (achieves Goal \#3).

\begin{algorithm}[t]
\begin{algorithmic}[1]
\begin{footnotesize}
\Function{SortCandidates}{$V$, $\zeta$} 

\Comment{$V$: VM to be placed, $\zeta$: list of candidate servers}

	\State $\omega \gets V^{PredictedWorkloadType}$
	
	\For {$c_i$ in $\zeta$}
		\State $\kappa_i \gets \Call{ScoreChassis}{c_i.Chassis}$
		\State $\eta_i \gets \Call{ScoreServer}{\omega, c_i}$
		\State $c_i.score \gets \alpha \times \kappa_i + (1 - \alpha) \times \eta_i $
	\EndFor

	\State \textbf{return} $\zeta.\Call{sortDesc}{c_i.score}$

\EndFunction

\Function{ScoreChassis}{$C$} 
	\For {$n_i$ in $C.Servers$}
		\For {$v_j$ in $n_i^{VMs}$}
			\State $\rho^{Peak} \gets \rho^{Peak} + v_j^{PredictedP95Util} \times v_j^{cores}$
		\EndFor
		\State $\rho^{Max} \gets \rho^{Max} + n_i^{cores}$
	\EndFor
	\State \Return $1 - \left [ \frac{\rho^{Peak}}{\rho^{Max}} \right ]$
\EndFunction

\Function{ScoreServer}{$\omega$, $N$} 
	\For {$v_i$ in $N^{UF\_VMs}$}
		\State $\gamma^{UF} \gets \gamma^{UF} + v_i^{PredictedP95Util} \times v_i^{cores}$
	\EndFor
	\For {$v_i$ in $N^{NUF\_VMs}$}
		\State $\gamma^{NUF} \gets \gamma^{NUF} + v_i^{PredictedP95Util} \times v_i^{cores}$
	\EndFor

	\If {$\omega = UF$}
		\State \Return $\frac{1}{2} \times \left( 1 + \frac{\gamma^{NUF} - \gamma^{UF}}{N^{cores}} \right)$
	\Else
		\State \Return $\frac{1}{2} \times \left( 1 + \frac{\gamma^{UF} - \gamma^{NUF}}{N^{cores}} \right)$
	\EndIf
\EndFunction
\end{footnotesize}
\end{algorithmic}
\caption{Criticality- \& utilization-aware VM placement}
\label{alg:workloadawareplacement}
\end{algorithm}

Algorithm~\ref{alg:workloadawareplacement} shows our rule
($SortCandidates$) and two supporting routines. We show the
predictions with the {\em PredictedWorkloadType} and {\em
  PredictedP95Util} superscripts. The rule ultimately computes the
score for each candidate server (line \#6). The higher the score, the
more preferable the server. The score is a function of how preferable
the server (line \#5) and its chassis (line \#4) are for the VM to be
placed. Both server and chassis intermediate scores range from 0 to
1. We weight the intermediate scores to give them differentiated
importance. We select the best value for the $\alpha$ weight in
Section~\ref{sec:scheduler}.

Function $ScoreChassis$ computes the chassis score for a candidate
server by conservatively estimating its aggregate chassis CPU
utilization, i.e. assuming all VMs scheduled to the chassis are at
their individual 95th-percentile utilization at the same time.  This
value is the sum of the predicted 95th-percentile
utilizations for the VMs scheduled to the chassis, divided by 
the maximum core utilization (\#cores in chassis$\times$100\%).
This ratio is proportional to utilization. We subtract it from 1, so
that higher values are better (line \#13).

Function $ScoreServer$ scores a candidate server differently depending
on the type of VM that is being deployed. First, it sums up the
predicted 95th-percentile utilizations of the user-facing VMs (lines
\#15-16) and non-user-facing VMs (lines \#17-18) independently. When a
user-facing VM is being deployed, we compute how much more utilized
the non-user-facing VMs on the server are than the user-facing
ones. We do the reverse for a non-user-facing VM.  The reversal is the
key to balancing the cap-able power across servers.  Adding 1 and
dividing by 2 ensure that the resulting score will be between 0 and 1
(lines \#20 and \#22), while higher values are better.

The algorithm takes only 7 milliseconds, which is negligible as
VM creation takes on the order of seconds~\cite{abrita2019benchmarking}.

\subsection{Per-VM power capping controller}
\label{sec:controller}

Azure's existing in-server capping system uses RAPL and
out-of-band (i.e., independently of software on the server) PSU alerts
to throttle all cores equally.  This approach is simple and
safe, but may strongly impact the user-facing workloads.  On the other
hand, a software-only approach that runs in-band (i.e., as a
platform-level service) can be aware of workload types, poll the PSUs,
and use per-core DVFS to cap non-user-facing VMs when necessary.
However, this solution is vulnerable to software defects and
communication issues.

To achieve safety and flexibility, we use a hybrid solution.  Our modified
chassis manager polls the PSUs every 200 milliseconds and alerts our
in-band controller when the chassis power draw is
close to the chassis budget.
The controller uses per-core DVFS to cap the cores running
non-user-facing VMs. To account for (1) high power draws that may
occur between polls or (2) the inability of the controller to bring
power below the budget, we use the out-of-band mechanisms as a
backup.

To power-manage the cores per-VM, we use the core-grouping feature of
the hypervisor (e.g., cpupools in Xen, cpugroups in Hyper-V) to split
the logical cores into two classes: high-priority and low-priority.
We assign the user-facing VMs and the I/O VM (e.g., Domain0 in Xen,
Root VM in Hyper-V) to the high-priority class, and the
non-user-facing VMs to the low-priority one.  The hypervisor ensures
that any threads associated with the given VM are scheduled only on
the logical cores of its group.

Upon receiving an alert from the chassis manager, the per-VM power
controller
compares the
server's power draw to its
budget. If the current draw is higher than the budget, the controller
immediately lowers the frequency of the low-priority cores to the
minimum p-state, i.e. half of the maximum frequency; the lowering of
the frequency may entail a lower voltage as well.  The goal is to
lower the server's power draw as quickly as possible without affecting
the important workloads. However, this large frequency reduction may
overshoot the needed power reduction. To reduce the impact on the
non-user-facing VMs, the controller then enters a feedback loop where each
iteration involves (1) checking the server power meter and (2)
increasing the frequency of $N$ low-priority cores to the next higher
p-state, until the power is close
to the budget. It selects the highest frequency that keeps the power
below this threshold. $N$ = 4 works well in our experiments.

It is possible that cutting the frequency of the low-priority cores in
half is not enough to bring the power below the server's budget.  For
example, a VM placement where there are not enough non-user-facing VMs
in the workload mix, non-user-facing VMs exhibiting lower utilization
than predicted, or a controller bug can cause this problem.
In this case, the out-of-band mechanism will kick in as backup. Though
RAPL will apply to all cores indiscriminately, protection from
overdraw must take precedence over performance loss.

The controller lifts the cap after some time (30 secs by default),
allowing all cores to return to maximum performance.

\subsection{Oversubscription strategy}
\label{sec:strategy}

We now describe our oversubscription strategy, which uses our capping
system and placement policy, historical VM arrivals, and historical
power draws, to increase server density.  It relies on the algorithm
below for computing an aggressive power budget for all the chassis of
each hardware generation.  Adapting it to find budgets for larger
aggregations (e.g., rack, row) is straightforward.  We refer to the
uncapped, nominal core frequency as the ``maximum'' frequency.

To configure the algorithm, we need to select the maximum acceptable
rate of capping events (e.g., \#events per week) for
user-facing ($emax_{UF}$) and non-user-facing ($emax_{NUF}$) VMs, and
the minimum acceptable core frequency (e.g., half the maximum
frequency) for user-facing ($fmin_{UF}$) and non-user-facing
($fmin_{NUF}$) VMs. If we want no performance impact for user-facing
VMs, we set $emax_{UF} = 0$ and $fmin_{UF} =$ maximum frequency.
As we describe next, our 5-step algorithm finds the lowest
chassis power budget that satisfies $emax_{UF}, emax_{NUF},
fmin_{UF},$ and $fmin_{NUF}$.

\noindent{\bf Estimate future behaviors based on history:}
\begin{enumerate}[wide,labelwidth=!,labelindent=10pt,topsep=0pt,itemsep=-1ex,partopsep=1ex,parsep=1ex]
\item Estimate the historical average ratio of user-facing virtual
  cores in the allocated cores ($\beta$).  Estimate the historical
  average P95 utilization of virtual cores in user-facing
  ($util_{UF}$) and non-user-facing ($util_{NUF}$) VMs.
\end{enumerate}

\noindent{\bf Profile the hardware:}
\begin{enumerate}[wide,labelwidth=!,labelindent=10pt,topsep=0pt,itemsep=-1ex,partopsep=1ex,parsep=1ex]
\setcounter{enumi}{1}
\item
  Estimate how much server power
  can be reduced by lowering core frequency
  at $util_{UF}$ and $util_{NUF}$, given $fmin_{UF}$ and
  $fmin_{NUF}$, respectively. This step produces two curves for power
  draw (one curve for each average utilization), as a function of
  frequency.
\end{enumerate}

\noindent{\bf Compute power budgets based on historical draws:}
\begin{enumerate}[wide,labelwidth=!,labelindent=10pt,topsep=0pt,itemsep=-1ex,partopsep=1ex,parsep=1ex]
\setcounter{enumi}{2}
\item Sort the historical chassis-level power draws
  (one reading per chassis per unit of time) in descending order.
\item Start from the highest power draw as the first candidate
  budget and progressively consider lower draws until we find
  $P_{min}$.  For each candidate power budget, we check that the rate
  of capping events would not exceed $fmax_{UF}$ or $fmax_{NUF}$
  (considering the higher draws already checked), and the attainable
  power reduction from capping is sufficient (given $\beta$ and the
  curves from step 2).
\item To compute the final budget,
  add a buffer (e.g., 10\%) to the budget from
  step 4 to account for future variability of $\beta$ or
  substantial increases in chassis utilization.
\end{enumerate}

We can use the difference between the overall budget computed in step
5 and the provisioned power to add more servers to the datacenter.
Because we protect user-facing VMs and use our VM scheduling policy,
this difference is substantially larger than in prior approaches, as
we show in Section~\ref{sec:eval-oversub}.

\noindent{\bf Example.} Suppose we are willing to accept rates of
0.1\% and 1\% capping events for user-facing and non-user-facing VMs,
respectively.  Suppose further that, upon a capping event, we are
willing to lower the core frequencies to 75\% and 50\% of the maximum,
respectively.
Now, assume we have 10000 historical chassis power draws
(collected from every chassis), and that the highest
draws have been 2900W, 2850W, and 2850W. We first consider 2900W. If
we were to set the chassis budget to just below that value to say
2890W, there would be 1 capping event out of 10000 observations,
i.e. a rate of 0.01\%, and we would have to
shave 10W during the event.  Given the acceptable capping rates and
minimum frequencies, we can operate with the data from step 1 and the
curves from step 2 to determine (1) whether we could reduce
power by 10W, and (2) whether there would be an impact on user-facing
VMs.
If we can achieve the reduction, we count 1 event out of 10000 that
would affect non-user-facing VMs. If the user-facing VMs would also
have to be throttled, we would count 1 event out of 10000 that would
affect those VMs.  Since both rates are lower than 0.1\% and 1\%, we
can now check a budget just below 2850W, say 2840W.  We repeat the
process for this budget, which would lead to 3 events out of 10000,
and the need to lower power by 60W (2900-2840) once and 10W
(2850-2840) twice.  Then the next lower draw and so on, until we
violate the desired capping rates
and minimum frequencies.
\section{Evaluation}
\label{sec:eval}

\subsection{Methodology}
\label{sec:method}

\noindent{\bf Data analysis.} We evaluate our criticality algorithm
and ML models (Section~\ref{sec:alg-models}) using standard metrics,
such as precision and recall from predictions.  We compute the metrics
based on the {\em entire VM workload of Azure in April 2019}.

\noindent{\bf Real experiments.} We run experiments on the same
hardware that Azure uses in production.  We use a chassis with
12 servers, each containing 40 cores split into two sockets.
At their nominal frequency, each server draws between 112W (idle) and
310W (100\% CPU utilization). At half this frequency, each server
draws from 111W to 169W.

Our single-server experiments (Section~\ref{sec:server}) explore the
dynamic behavior of our per-VM power capping controller and resulting
VM workload performance for a combination of user-facing and
non-user-facing VMs and various power budgets. For comparison, we use
the existing full-server capping controller (Intel's RAPL) in
Azure.
Our chassis-level experiments (Section~\ref{sec:chassis}) explore our
system on 12 servers, including PSU alerts.
For comparison, we use the existing
chassis-level mechanisms in Azure
(Figure~\ref{fig:chassispowercap}).

For both sets of experiments, we use instances of a latency-critical
transaction processing application (similar to TPC-E) for the
user-facing workload, and instances of a batch Hadoop computation
(Terasort) for the non-user-facing workload.  For the user-facing
workload, we use real inputs from Azure's team responsible for
it, whereas we use synthetic input data for the non-user-facing
computation.

\noindent{\bf Simulation.} We evaluate our modified VM scheduler in
simulation (Section~\ref{sec:scheduler}), leveraging the same
simulator that Azure uses to evaluate changes to the VM
scheduler before putting them in production; our only extension is to
simulate calls to the ML system.
An event generator drives the simulation with a sequence of VM
arrivals.  For each arrival, it invokes the production scheduling
algorithm (or the scheduling algorithm with the addition of our
policy) for server placement decisions.
Running the actual scheduler code in the simulator ensures that
simulations are faithful to reality.

We simulate a cluster of 60 chassis in 20 racks.
The simulator produces VM arrivals
based on distributions
matching Azure's load in April 2019.
Table~\ref{table:params} lists the main statistics.

\begin{table}[t]
\centering
{\footnotesize
\begin{tabular}{|p{0.39\columnwidth}|p{0.53\columnwidth}|}
\hline
\thead{Parameter} & \thead{Value}\\
\hline
Cluster configuration & 20 racks $\times$ 3 chassis $\times$ 12 blades\\
\hline
Blade configuration & 2x20 cores\\
\hline
VM size dist. (cores) & 1 (33\%), 2 (27\%), 4 (21\%), 8 (10\%), 16 (5\%), 24 (3\%), $>=$32 (1\%) \\
\hline
Deployment size dist. (\#VMs) & 1 (39\%), 2 (14\%), 3-5 (16\%), 6-10 (9\%), 11-15 (8\%), 16-25 (5\%), $>$25(9\%)\\
\hline
VM lifetime dist. (hours) & 1 (52\%), 2 (5\%), 3-5 (10\%), 6-10 (9\%), 10-25(7\%), 26-720 (8\%), $>$720 (9\%) \\
\hline
Workload type buckets & user-facing (UF), non-user-facing (NUF)\\
\hline
P95 utilization buckets & 0-25\%, 26-50\%, 51-75\%, 76-100\%\\
\hline
Avg UF:NUF core ratio & 4:6\\
\hline
Avg UF and NUF P95 util & 65\% (bucket \#3), 44\% (bucket \#2)\\
\hline
\# simulation days & 30\\
\hline
\end{tabular}
}
\vspace{-.05in}
\caption{Simulation parameters.}
\label{table:params}
\vspace{-.2in}
\end{table}

\begin{table*}
\parbox{.372\linewidth}{
\footnotesize{
\begin{tabular}{|c|c|c|c|}
	\hline
	Technique & Recall & Recall   & Precision \\
        & target & achieved & achieved \\
	\hline
	Pattern-matching & 99\% & 99\% & 76\% \\
	ACF     & 99\% & 99\% & 54\% \\
	FFT     & 99\% & 99\% & 48\% \\
	\hline
	Pattern-matching & 98\% & 98\% & 77\% \\
	ACF     & 98\% & 98\% & 56\% \\
	FFT     & 98\% & 98\% & 50\% \\
	\hline
\end{tabular}
\caption{Pattern-matching vs ACF vs FFT.}
\label{tab:algs}
}
}
\hspace{-.03in}
\parbox{.62\linewidth}{
\footnotesize{
\begin{tabular}{|c|c|c|c|c|c|c|c|}
\hline	
Prediction & Model & \% High & Bucket 1 & Bucket 2 & Bucket 3 & Bucket 4 & Accuracy \\
           &       & Conf.   & R | P    & R | P    & R | P    & R | P    & \\
\hline
Criticality & GB & 99\% & 67\% | 77\% & 99\% | 99\% & NA & NA & 98\% \\
            & RF & 99\% & 69\% | 78\% & 99\% | 99\% & NA & NA & 98\% \\
\hline
P95 util & GB & 68\% & 95\% | 85\% & 47\% | 77\% & 51\% | 79\% & 94\% | 80\% & 82\% \\
         & RF & 73\% & 93\% | 87\% & 61\% | 76\% & 65\% | 81\% & 92\% | 83\% & 84\% \\
\hline
\end{tabular}
\caption{Random Forest (RF) and Gradient Boosting (GB) models recall
  (R), precision (P), and accuracy for high-confidence predictions.}
\label{tab:models}
}
}
\vspace*{-.2in}
\end{table*}

\begin{figure}[t]
\centering
\includegraphics[height=2in,width=0.8\columnwidth]{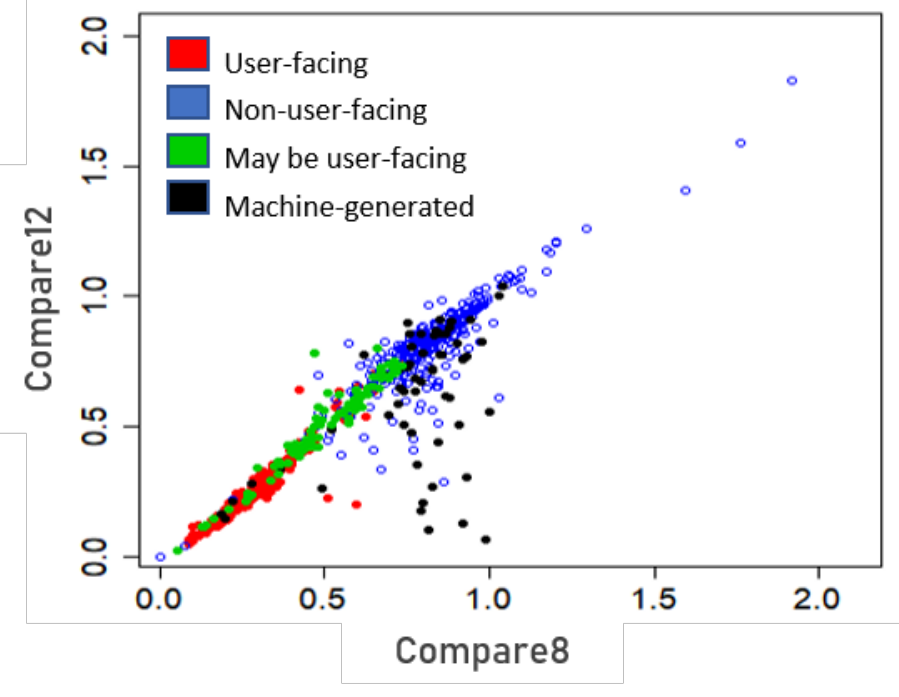}
\vspace{-.15in}
\caption{Algorithm compared to manual classification.}
\label{fig:hand-class}
\vspace{-.25in}
\end{figure}

\subsection{Criticality algorithm and ML models}
\label{sec:alg-models}

\noindent{\bf Criticality algorithm.}  To evaluate our
pattern-matching algorithm (Section~\ref{sec:ML}), we first compare
its classifications to our own manual labeling of 840 workloads on
Azure.  Figure~\ref{fig:hand-class} shows one colored dot for each
workload with coordinates corresponding to its Compare8 and Compare12
values.  The colors indicate whether we deem the workload clearly
user-facing, possibly user-facing, clearly machine-generated, or
clearly non-user-facing.  The figure shows that Compare8 can separate
the first two groups, which the algorithm should conservatively
classify as user-facing, from the last two.  A vertical bar at
Compare8=0.72 gets all important workloads to the left of the bar, and
the vast majority of unimportant ones to the right of it.  Compare12
does not separate the classes well.

Thus,
the algorithm can accurately classify
workloads based on their Compare8 value.  For a quantitative
assessment, we compare it to two well-known approaches for finding
periodicity in a time series, ACFs and FFTs, for the same set of
workloads. For both approaches, we do the same
pre-processing
and disambiguate between user-facing and machine-generated workloads
using the same methods as in our algorithm.

As we want to protect user-facing VMs, we must achieve high
  recall for this class as the recall indicates the probability of
  correctly identifying these VMs.  Table~\ref{tab:algs} shows the
precision and recall, for two high recall targets (0.99 and 0.98) for
whether a workload is user-facing.  Our algorithm achieves the target
recall with much higher precision than its counterparts.

\noindent{\bf ML models.}  We now evaluate our models for {\em
  Azure's entire VM workload}.  Table~\ref{tab:models} lists the
percentage of predictions with confidence score higher than 60\% (3rd
column), and the per-bucket recalls and precisions (4th-7th columns)
and the accuracy (rightmost column) for those high-confidence
predictions.  The VM scheduler disregards predictions with lower
confidence and conservatively assumes the VM being deployed will be
user-facing and will exhibit 100\% 95th-percentile utilization.  For
comparison, we show results for the equivalent Gradient Boosting (GB)
models.

The table shows that our criticality model achieves 99\% recall for
user-facing VMs (Bucket 2), which is critical for protecting these
VMs. The most important features for our model are the percentage of
user-facing VMs observed in the cloud subscription, the percentage of
VMs that live longer than 7 days in the subscription, and the total
number of VMs in the subscription. The GB model achieves similar
results.

Our utilization model also does well with good recall and precision
(83-93\%) for the most popular buckets (1 and 4), and good accuracy
(84\%) for the 73\% of high-confidence predictions. Here, the most
important features are the average of the VMs' 95-percentile CPU
utilizations in the subscription, the average of the VMs' average CPU
utilizations in the subscription, and the percentages of VMs in each
CPU utilization bucket in the subscription. The GB model achieves
similar accuracy, but with fewer high-confidence predictions and
lower recall for the two middle (least popular) buckets.

\subsection{Per-VM capping controller experiments}
\label{sec:server}

We run experiments on a server with our user-facing appli\-cation
running on a VM with 20 virtual cores and our non-user-facing
application running simultaneously on another VM with 20 virtual
cores.  Each execution takes 10 minutes.

\begin{figure}[t]
\centering
\includegraphics[width=0.95\columnwidth]{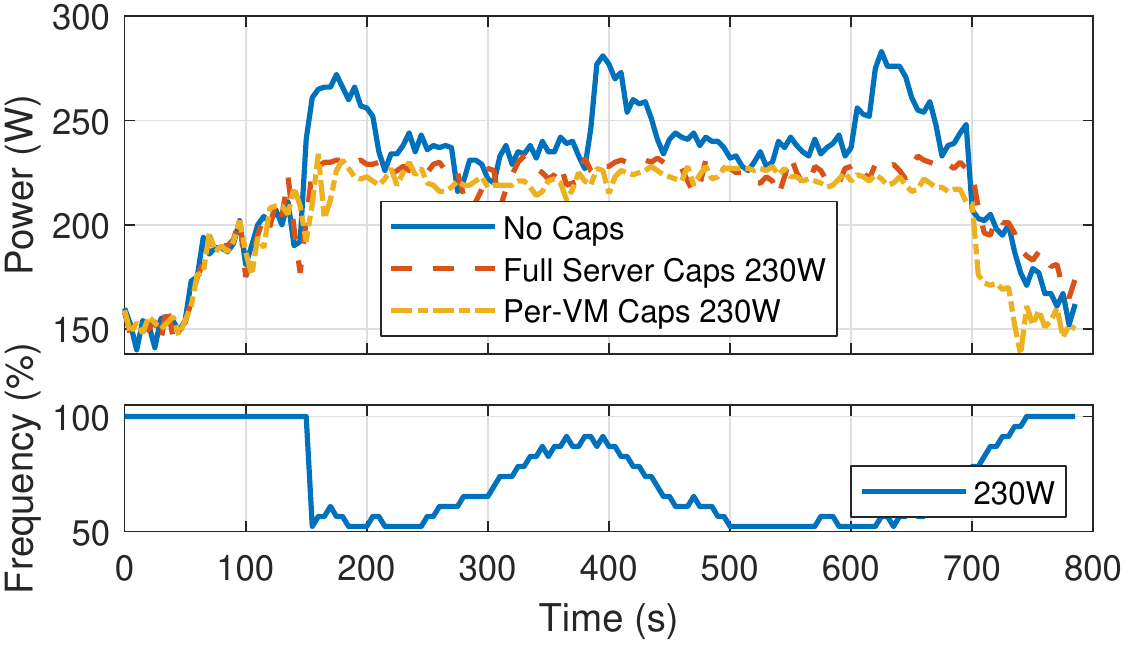}
\vspace{-.1in}
\caption{Server power dynamics.}
\label{fig:dynamics}
\vspace{-.2in}
\end{figure}

Figure~\ref{fig:dynamics} plots the dynamic power behaviors and core
frequencies of full-server and per-VM capping with caps at 230W.  In
the bottom graph, we plot the lowest frequency of any non-user-facing
core.  The experiments have capping enabled throughout their
executions. For comparison, we show the power profile of an experiment
without any cap.

When unconstrained (no cap), the power significantly exceeds
250W. In contrast, full-server and per-VM
capping keep the power draw below 230W. Because of the lower target of
our controller (225W for the 230W cap), its power draws are slightly
below those of full-server capping most of the time. The 
frequency curve depicts the adjustments that our controller makes
to the performance of the non-user-facing VM. The steep drop to the
lowest frequency occurs when the controller abruptly lowers the
frequency to the minimum value when the power first exceeds the
target. After that point, its feedback component smoothly increases
and decreases the frequency.

\begin{figure}[t]
\centering
\includegraphics[height=1.5in,width=0.95\columnwidth]{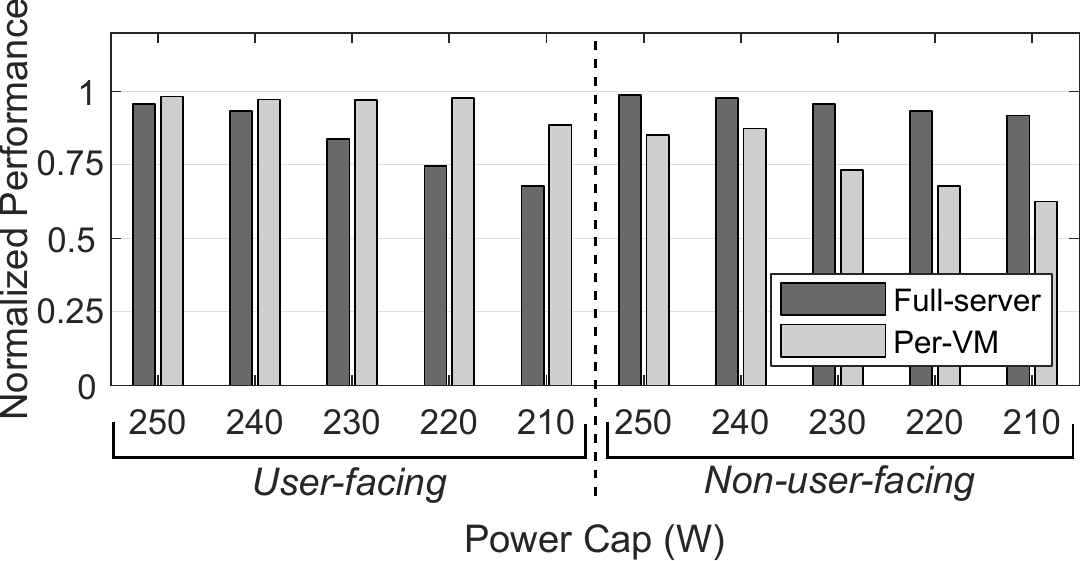}
\vspace{-.1in}
\caption{Performance impact of power capping.}
\label{fig:impact}
\vspace{-.21in}
\end{figure}

Figure~\ref{fig:impact} shows the impact of capping during these
experiments on the 95th-percentile latency of the user-facing
application (10 leftmost bars) and the running time of the
non-user-facing application (10 rightmost bars). Results are
normalized to the unconstrained performance of each application. We
also include bars for capping at 250W, 240W, 220W, and 210W.

The results show that full-server capping imposes a large tail latency
degradation, especially for the lower caps.  When the cap is 230W, the
degradation is already 18\%, which is often unacceptable for
user-facing applications.  For lower caps, full-server capping
provides even worse tail latency (35\% degradation for 210W).
In contrast, our controller keeps tail latency very close to the
unconstrained case, until the cap is so low (210W) that it becomes
impossible to protect the user-facing application and RAPL needs to
engage. This positive result comes at the cost of performance loss for
the non-user-facing application. While full-server capping keeps
running time fairly close to the unconstrained case, our controller
degrades it by 28\% for the 230W cap. This is the right tradeoff, as
non-user-facing workloads have looser performance needs.

\subsection{Chassis-level capping experiments}
\label{sec:chassis}

We now study the power draw at the chassis level, and the
impact of different capping granularities (full-server vs
per-VM) and VM placements.  We experiment with a 12-server chassis
running 36 copies of our user-facing application (each on a VM with 4
virtual cores), and 36 copies of our non-user-facing application (each
running on a VM with 6 virtual cores).
In terms of VM placement, we explore two extremes: (1) {\em balanced}
placement, where we place the user-facing and non-user-facing VMs in
round-robin fashion across the servers, i.e. 3 VMs of each type on
each server; and (2) {\em imbalanced}, where we segregate user-facing
and non-user-facing VMs on different sets of servers.  Each experiment
runs for 26 minutes.

\begin{figure}[t]
\centering
\includegraphics[width=\columnwidth]{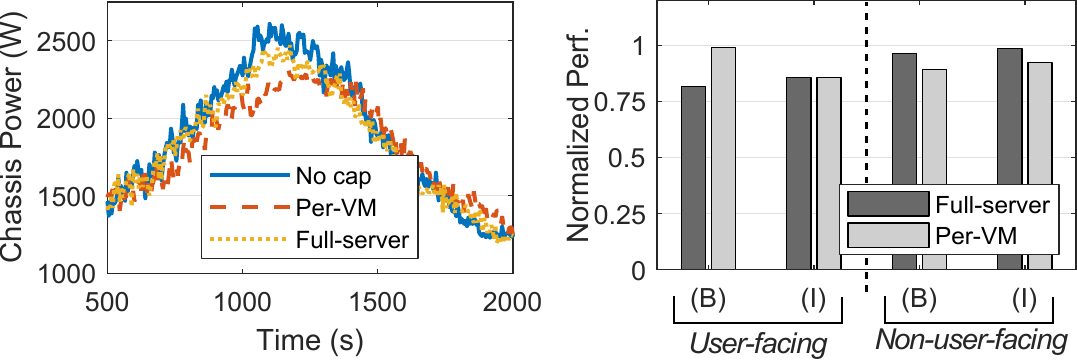}
\vspace{-.25in}
\caption{Chassis dynamics and performance vs placement.}
\label{fig:impact-chassis}
\vspace{-.2in}
\end{figure}

Figure~\ref{fig:impact-chassis}(left) plots the dynamic behavior of
the chassis for an overall budget of 2450W for the two capping
approaches.  For comparison, we also plot the no-cap case.  In these
experiments, we use the balanced placement as an example.

As expected, both capping granularities are able to limit the power
draw to the chassis budget, whereas the no-cap experiment
substantially exceeds this value.  We observe the same trends under
the imbalanced placement approach.  The VM placement does not matter
in terms of the power profiles because the capping enforcement ensures
no budget violations.

However, VM placement has a large impact on application
performance. Figure~\ref{fig:impact-chassis}(right) plots the impact
of VM placement and capping granularity on the average 95th-percentile
latency of the user-facing applications (4 leftmost bars) and on the
average running time of the non-user-facing applications (4 rightmost
bars).  We normalize to the no-cap results.

Per-VM capping under a balanced placement keeps the average tail
latency the same as the no-cap experiment, despite the tight 2450W
budget. In contrast, per-VM capping degrades performance as much as
full-server capping when the placement is imbalanced.  These results
show that our controller protects user-facing VMs when the VM
placement allows it.

\begin{figure*}[t]
    \centering
	\includegraphics[width=\textwidth]{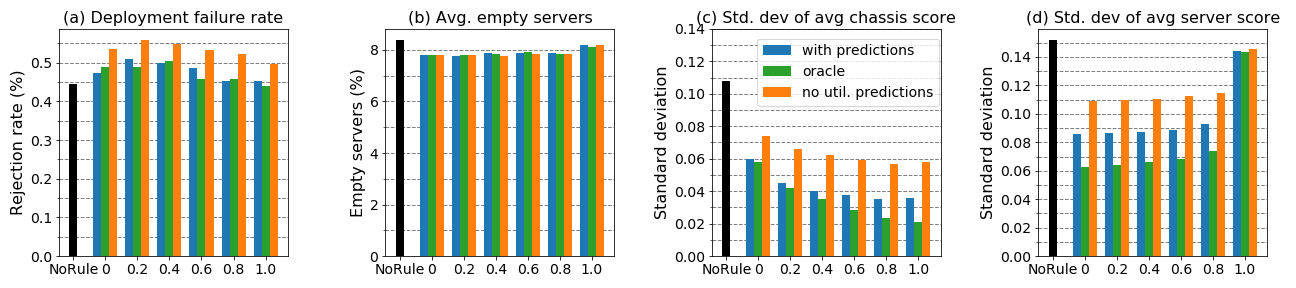}
     \vspace{-.25in}
     \caption{Key scheduler metrics, as a function of $\alpha$.}
     \label{fig:default}
\vspace*{-.2in}
\end{figure*}

Full-server capping provides slightly better performance than per-VM
capping for the non-user-facing applications.  More interestingly, the
results for balanced placement are slightly worse than for imbalanced
placement, regardless of the capping granularity.  For per-VM capping,
the reason is that servers with only non-user-facing VMs need to
reduce the frequency of fewer cores. For full-server capping, the
reason is that servers with only non-user-facing VMs tend to have
higher utilization, so a smaller reduction in frequency is enough for
a large power reduction. Comparing the two rightmost bars, we see that
full-server capping hurts performance slightly less than per-VM
capping in the imbalanced case, as RAPL lowers frequency more slowly
than our controller.

\subsection{Cluster VM scheduler simulation}
\label{sec:scheduler}

In the previous section, we explored extreme and manually-produced VM
placements in controlled experiments.  However, in practice,
placements are determined by the VM scheduler.  To evaluate our
modified scheduler, we implement our VM placement policy in
Azure's VM scheduler, and simulate a cluster using 30 days of VM
arrivals (Section~\ref{sec:method}). The simulator reports four main
metrics:
\begin{itemize}[wide,labelwidth=!,labelindent=10pt,topsep=0pt,itemsep=-1ex,partopsep=1ex,parsep=1ex]
\item Deployment failure rate: the percent of VM deployment requests
  rejected by the scheduler due to resource unavailability or
  fragmentation. This rate has a direct impact on users, so our policy
  should not increase it;
\item Average empty server ratio: the percent of servers without any
  VMs in the cluster averaged over time. Empty servers can accommodate
  the largest VM sizes, so our policy should not decrease this ratio;
\item Standard deviation of the average chassis score, i.e. $1 -
  (\rho^{Peak}/\rho^{Max})$
  (Algorithm~\ref{alg:workloadawareplacement}, line 13), for each
  chassis. This metric reflects how balanced the chassis are with
  respect to their power loads. Lower values are better and mean
  better balance and fewer power capping events;
\item Standard deviation of the average server score, i.e. $(1/2)
  \times (1 + (\gamma^{NUF} - \gamma^{UF})/N^{cores})$
  (Algorithm~\ref{alg:workloadawareplacement}, line 20), for each
  server. This metric reflects how balanced the servers are in terms
  of UF and NUF core 95th-percentile utilizations. Lower values mean
  better balance and that we are more likely to only need to cap NUF
  VMs.
\end{itemize}

\noindent{\bf Results.} Figure~\ref{fig:default} shows the results for
these metrics, as a function of the $\alpha$ weight in our policy
(Algorithm~\ref{alg:workloadawareplacement}, line 6). $\alpha = 1$
means that the server score is irrelevant, whereas $\alpha = 0$ means
that the chassis score is irrelevant. From left to right in each
graph, the ``NoRule'' (black) bar represents the existing scheduler;
the leftmost (blue) bar in each group represents our modified
scheduler using our policy and ML predictions; the next (green) bar
represents the modified scheduler using oracle predictions; and the
rightmost (orange) the modified scheduler using criticality
predictions, but no utilization predictions.

Comparing the black and blue bars illustrates the benefit of our
policy and predictions.  Figures~\ref{fig:default}(a) and (b) show
that our modified scheduler impacts the failure rate slightly for low
values of $\alpha$ (not at all for high values), while slightly
decreasing the percentage of empty servers regardless of $\alpha$.
The reason is that our policy may use a few more servers in the
interest of better balancing the load.  In fact,
Figures~\ref{fig:default}(c) and (d) confirm that the load is more
balanced using our policy and predictions. These latter graphs also
show that the value of $\alpha$ is important. $\alpha = 0$ produces
much worse utilization balancing across chassis than other values.  At
the same time, $\alpha = 1$ produces as poor server utilization
balancing as the existing scheduler, whereas other values produce much
better server balancing.  These observations confirm that it is key to
balance both across chassis and servers, as in our policy. $\alpha =
0.8$ strikes a good compromise between the importance of these types
of balancing.

\noindent{\bf Impact of prediction accuracy.}  Comparing the blue and
green bars illustrates the impact of mispredictions and predictions
with low confidence. Figures~\ref{fig:default}(c) and (d) show that
oracle predictions produce only slightly better balancing than our
real predictions for certain values of $\alpha$.

\noindent{\bf Impact of utilization predictions.} Comparing the blue
and orange bars illustrates the importance of having both criticality
and utilization predictions.  Clearly, it is critical to predict the
workload type of each VM, as we want to protect the performance of
user-facing VMs during capping events.
The results demonstrate that having utilization predictions is also
important.  The lack of such predictions degrades the balancing
substantially for most values of $\alpha$, and thus increases the
capping rate and limits the power reduction during an event (thereby
decreasing the potential for oversubscription).

\subsection{Oversubscription increases}
\label{sec:eval-oversub}

We now estimate the amount of oversubscription and dollar savings that
result from our lower chassis power budgets.  To do so, we translate
the amount of budget we can reduce in each 60-chassis cluster into the
infrastructure cost we would avoid.  Oversubscription allows servers
to be added to an existing datacenter (assuming space, cooling, and
networking are available, as it is often the case), avoiding the cost
of building a corresponding fraction of a new datacenter.

We instantiate our 5-step oversubscription strategy
(Section~\ref{sec:strategy}) with power telemetry from 1440
chassis over 3 months in 2018. We also use VM statistics
from April 2019.  Specifically, the 95th-percentile core utilizations
for non-user-facing ($util_{NUF}$) and user-facing ($util_{UF}$) VMs
were 44\% and 65\%, respectively, and the ratio of user-facing cores
in the allocated cores ($\beta$) was 40\%. We add a buffer of 10\%
to the chassis budget (step 5).
For the results where providers treat external (i.e., third-party) VMs
differently than internal (i.e., first-party) VMs, we adjust these
parameters accordingly, while keeping the same amount of buffer.

Table~\ref{tab:oversubsim} lists results for several types of
provisioning:
\begin{enumerate}[wide,labelwidth=!,labelindent=10pt,topsep=0pt,itemsep=-1ex,partopsep=1ex,parsep=1ex]
  \item ``Traditional'' provisioning (no oversubscription);
  \item State-of-the-art full-server capping without VM insights. In
    this approach, the power capping events need to be rare and the
    throttling has to be light to prevent performance loss to
    user-facing VMs.  To model this approach with our provisioning
    strategy, we use $emax_{UF} + emax_{NUF} =$ 0.1\%, $fmin_{UF} =
    fmin_{NUF} =$ 75\%;
  \item Predictions-based per-VM capping and scheduling, without
    impact on user-facing VMs. We use $emax_{UF}$ = 0,
    $fmin_{UF}$ = 100\%, $emax_{NUF}$ = 1\% and $fmin_{NUF}$ = 50\%;
  \item Predictions-based per-VM capping and scheduling, with minimal
    impact on user-facing VMs. To make this approach
    comparable to the others, we set the overall rate of capping
    events at 1\%. Specifically, we use $emax_{UF}$ = 0.1\%,
    $fmin_{UF}$ = 75\%, $emax_{NUF}$ = 0.9\% and $fmin_{NUF}$ = 50\%;
  \item Predictions-based per-VM capping and scheduling for internal
    VMs only (all external VMs considered user-facing), without
    impact on user-facing VMs;
  \item Predictions-based per-VM capping and scheduling for internal
    VMs only (all external VMs considered user-facing), with minimal
    impact on user-facing VMs;
  \item Predictions-based per-VM capping and scheduling for internal
    and non-premium external VMs, without impact on
    user-facing VMs; and
  \item Predictions-based per-VM capping and scheduling for internal
    and non-premium external VMs, with minimal impact on
    user-facing VMs.
\end{enumerate}

\begin{table}[t]
  \begin{center}
    {\footnotesize
    \begin{tabular}{c|c|c} 
      Approach & Chassis budget & Savings\\ 
      ~        & delta (\%)     & (\$10/W)\\ 
      \hline
      Traditional & 0 & 0\\
      State of the art & 6.2\% & \$79.4M\\
\hline
      Predictions for all VMs,       &        &         \\
      no UF impact                   & 11.0\% & \$140.8M\\
      Predictions for all VMs,       &        &         \\
      minimal UF impact              & 12.1\% & \$154.9M\\
\hline
      Predictions for internal VMs,  &        &         \\
      no UF impact                   & 8.4\%  & \$107.5M\\
      Predictions for internal VMs,  &        &         \\
      minimal UF impact              & 10.3\% & \$131.8M\\
      Predictions for internal and   &        &         \\
      non-premium external VMs,      &        &         \\
      no UF impact                   & 10.6\% & \$135.7M\\
      Predictions for internal and   &        &         \\
      non-premium external VMs,      &        &         \\
      minimal UF impact              & 12.1\% & \$154.9M\\
\hline
    \end{tabular}
    }
    \caption{Comparison between provisioning approaches.}
    \label{tab:oversubsim}
  \end{center}
  \vspace{-.3in}
\end{table}

The state-of-the-art approach achieves 6.2\%
oversubscription. This amount is comparable to that (8\%) achieved by
Facebook~\cite{Wu2016}, which knows the (single) workload that runs on
each server. Public cloud platforms do not have this luxury.

In contrast, our approach (\#3 and \#4) can {\em almost double} the
oversubscription and savings.  We achieve more than 12\%
oversubscription with minimal impact on user-facing VMs.  Assuming a
datacenter campus of 128MW and an infrastructure cost of
\$10/W~\cite{barroso2018datacenter}, 12.1\% oversubscription
translates into \$154.9M in savings; an increase in savings of \$75.5M
over the state of the art.
As providers can oversubscribe many campuses, the savings would be
much higher in practice.

When providers prefer to treat external and internal VMs differently
(approaches \#5-\#8), they can do so at the cost of a lower increase
in oversubscription.  For example, when treating all external VMs as
user-facing and protecting the performance of user-facing VMs (\#5),
the increase in savings becomes \$28.1M.  At the other extreme, where
we treat the premium external VMs as user-facing and allow minimal
impact on user-facing VMs (\#8), the increase in savings returns to
\$75.5M.  The reason is that this provisioning approach has enough
non-critical VMs, and oversubscription is limited only by the rate of
capping events.

\section{Lessons from production deployment}
\label{sec:lessons}

So far, we have deployed our per-VM capping controller and ML models
in production in Azure's datacenters. Next, we discuss some of
the lessons from these deployments.

\noindent{\bf Hypervisor support for per-VM power capping.} Our
prototype controller (Section~\ref{sec:controller}) leveraged the
hypervisor's core-grouping feature to manage the frequency of each
VM's physical cores. In production, Azure typically prefers not
to restrict a VM to a subset of cores, so we could not rely on this
feature.  Instead, we had to extend the hypervisor to (1) add the
capability to dynamically specify the frequency for a VM, and (2)
carry the frequency to whichever cores it schedules the VM on during
the context switch (changing the frequency takes tens of microseconds,
whereas a scheduling quantum lasts 10 milliseconds).  As most VMs are
small (Table~\ref{table:params}), there was no need to manage
frequency on a per-virtual-core basis.

\noindent{\bf Additional types of throttleable VMs.} Some first-party
customers were concerned about the impact of per-VM capping on their
non-user-facing VMs. To alleviate their concerns, we added a
configurable prioritized throttling list to our system.  Using the
list, we first consider low priority and internal non-production VMs
for throttling and throttle production (including third-party, if
configured) non-user-facing VMs as a last resort, i.e. when throttling
the other types is insufficient.

\noindent{\bf Metrics to measure impact.} Since VMs are black boxes,
we cannot use any workload-specific metric to evaluate per-VM capping
in production.  Instead, our deployed system measures how long and how
hard VMs are being capped.  The data shows that our system is
successful at protecting production user-facing VMs, while
prioritizing the VMs that do get throttled.

\noindent{\bf Increasing rack density with per-VM capping.}  While
deploying our system, we learned that Azure was installing
fewer servers per rack -- 28 instead of 36 -- when deploying a new
generation of power hungrier servers.  Having fewer servers per rack
reduces the probability that the rack power draw will hit the
provisioned limit and cause capping using RAPL. With our per-VM
capping system in place, Azure will go back to installing 36
servers per rack. {\em This is another type of power oversubscription
  that per-VM capping enables.}

\noindent{\bf Killing VMs.} Some first-party customers indicated that
they would prefer their VMs to be killed rather than throttled, as
their services can handle losing VMs but an unpredictable impact due
to throttling is not acceptable.  Under extreme power draws, killing
these customers' VMs can help protect production user-facing VMs and
throttle fewer non-user-facing ones.  We will soon add this capability
to our system.

\noindent{\bf Server support for per-VM management.} Our production
experience has highlighted the drawbacks of managing VM power
per-component (e.g., core, uncore, memory).  We expect that cloud
providers would prefer to raise the level of abstraction from
individual components to entire VMs, even if VM power would have to
approximate.  This would enable advances that have been too complex
for production use, such as power-aware VM placement, enforcing per-VM
power limits, and making capping and killing decisions based on VM
power.  We are working with silicon vendors towards this end.
\section{Related Work}
\label{sec:related}

Our paper is {\em the first to use ML predictions for increasing power
  oversubscription in public cloud platforms.}  Next, we discuss some
of the most closely related works.

\noindent{\bf Leveraging predictions.} Some works predict resource
demand, resource utilization, or job/task length for provisioning or
scheduling purposes,
e.g.~\cite{Calheiros15,Cortez2017,Gong10,Islam12,Khan12,Roy11}. In
contrast, we introduce a new algorithm and ML models for predicting
workload type and high-percentile utilization, seeking to protect
critical workloads from capping and place VMs in a criticality- and
capping-aware manner.

\noindent{\bf Server power capping.}
Most efforts have focused on selecting the DVFS
setting
required to meet a tight power budget as applications execute,
e.g.~\cite{Guliani19,Isci06,Lefurgy07,Liu16,Lo15,Ma11,Mishra10,
  Raghavendra2008,Zhang16}.
Both modeling/optimization and feedback techniques have been used.
The inputs to the selection have been either application-level metrics
(e.g., request latency), low-level performance counters,
or operator annotations (e.g., high priority
application).  Our capping controller uses per-core DVFS and
feedback, so it adds to this body of work.  However, it
also uses predictions about the VMs' performance-criticality as its
inputs.
Our approach seems to be the best for a cloud platform, since criticality
information and application-level metrics are typically not available,
and collecting and tracking low-level counters at scale involves
undesirable overhead.

\noindent{\bf Cluster-wide workload placement/scheduling.}
Many works select workload
placements to reduce performance interference or energy usage,
e.g.~\cite{Beloglazov10,Bobroff07,Delimitrou2014,Dejan2013,Verma08,
  Yang2013}. Unfortunately, they are often impractical for a cloud
provider, relying on extensive profiling, application-level
metrics, short-term load predictions, and/or aggressive resource
reallocation (e.g., via live migration).  Live migration is
particularly problematic, as it retains contended resources, may
produce traffic bursts, and may impact VM availability; it is better
to place VMs where they can stay.
Our scheduler uses predictions in VM placement.  Unlike prior work, it
reduces the number and impact of capping events, and increases power
oversubscription.

\noindent{\bf Datacenter oversubscription.}
Researchers have proposed to use statistical oversubscription, where
one profiles the aggregate power draw of multiple services and deploy
them to prevent correlated
peaks~\cite{Fan2007,govindan2009,Hsu2018,Wang2016}.  Our work extends
these works by using predictions to place the workload, inform
capping, and increase oversubscription.  Our oversubscription strategy
is also the first to carefully control the extent and impact of
capping on important cloud workloads.

Others have studied hierarchical capping in production
datacenters~\cite{Fan2007,Hsu2018,Wu2016}.
Our paper focuses on chassis-level power budget enforcement to make
our experimentation easier.  However, our techniques extrapolate
directly.  For example, for row-level budget enforcement, we can place
VMs across rows trying to balance rows and servers.

Finally, researchers have proposed using energy storage to shave power
peaks in oversubscribed datacenters~\cite{Govindan2012,Kontorinis12}.
When peaks last long, this approach may require large amounts of
storage, which our work does not require.  Nevertheless, the two
approaches are orthogonal and can be combined.
\section{Conclusions}

We proposed prediction-based techniques for increasing power
oversubscription in cloud platforms, while protecting important
workloads.
Our techniques can increase oversubscription by $2\times$.
We discussed lessons from deploying our techniques in production.  We
conclude that recent advances in ML and prediction-serving systems can
unleash further innovations in cloud resource provisioning and
management.

%%%%%%% -- PAPER CONTENT ENDS -- %%%%%%%%

%%%%%%%%% -- BIB STYLE AND FILE -- %%%%%%%%
\bibliographystyle{IEEEtranS}
\bibliography{references}
%%%%%%%%%%%%%%%%%%%%%%%%%%%%%%%%%%%%

\end{document}